\documentclass[twocolumn,aps,superscriptaddress,nolongbibliography,pra,10pt]{revtex4-2}
\usepackage[utf8]{inputenc}
\usepackage{graphicx}
\usepackage{amsmath}
\usepackage{amssymb}
\usepackage{bm}
\usepackage{hyperref}
\usepackage{verbatim}

\usepackage{xcolor}

\begin{document}

\title{Hyperfine van der Waals repulsion between open-shell polar molecules}
\author{Etienne F. Walraven}
\affiliation{Institute for Molecules and Materials, Radboud University, Nijmegen, The Netherlands}
\author{Tijs Karman}
\email{tkarman@science.ru.nl}
\affiliation{Institute for Molecules and Materials, Radboud University, Nijmegen, The Netherlands}

\begin{abstract}
   We describe a novel type of interaction between open-shell polar molecules at sub-millikelvin temperatures. This hyperfine van der Waals interaction occurs between two molecules in two rotational states that differ by one quantum. Normally, this induces resonant dipolar interactions that lead to rapid collisional loss. For specific hyperfine states, however, selection rules prevent this. One can effectively turn off the dipolar interaction by merely flipping a nuclear spin. The resulting van der Waals interaction can be repulsive and can suppress collisional loss. We focus on laser-coolable CaF, but show this effect occurs universally for open-shell molecules, including MgF, SrF, BaF and YO. We propose that this effect could be measured by merging molecules in optical tweezers, where flipping a spin in one of the tweezers enables tuning of collision rates by five orders of magnitude.
\end{abstract}

\maketitle

\section{Introduction \label{sec:intro}}

Ultracold dipolar molecules are a promising platform for quantum simulation \cite{micheli:06,yan:13,blackmore:18ref,altman:21} and quantum information \cite{demille:02,kaufman:21,yelin:06,ni:18ref,holland:23,bao:23}. A crucial requirement is the ability to control their long-range interactions. The use of external fields has led to precise control over molecular collisions by controlling the interaction potentials between collision partners \cite{langen:24,cornish:24,karman:24}. For example, engineering repulsive interactions through shielding with microwaves \cite{karman:18d,karman:19c,karman:20,anderegg:21,schindewolf:22} or static electric fields \cite{quemener:16ref,gonzalez:17ref,matsuda:20,li:21,mukherjee:23} suppresses collisional loss \cite{takekoshi:14,voges:20,guo:18,ye:18,gregory:19,cheuk:20ref} and has enabled the cooling of molecular gases to quantum degeneracy \cite{valtolina:20,matsuda:20,schindewolf:22,bigagli:24}. Interactions between atoms can even be controlled without external fields by preparing them in Rydberg states \cite{raimond:81,tong:04,urban:09,shaffer:18}. In this paper, we explore how preparing molecules in selected hyperfine states in the electronic ground state can give rise to resonant dipole-dipole and van der Waals interactions.

The dipole-dipole interaction is the dominant long-range interaction between polar molecules \cite{stone:96,gray:76ref}. However, for molecules in a single quantum state, their dipole moments average to zero in the absence of external fields. Consequently, the first-order interaction vanishes and the interaction is dominated by dipole-dipole in second order, commonly referred to as the van der Waals interaction.
For many molecules, this interaction is dominated by coupling to electronically excited states, which leads to attractive interactions.
However, virtual excitations to higher rotational states within the electronic ground state can also give rise to van der Waals interactions. For many dipolar molecules used in ultracold physics, this rotational contribution exceeds the electronic contribution by two orders of magnitude \cite{zuchowski:13}, as rotational excitation energies can be five orders of magnitude smaller than electronic ones. We recently found that while two molecules in the same rotational state experience an attractive rotational van der Waals interaction, two molecules in different rotational states that differ by more than one quantum experience a repulsive interaction. This repulsion is a result of dipole-dipole coupling to a lower-lying pair of rotational states \cite{walraven:24a}.

There is one exception where the dominant interaction is not a van der Waals interaction, but rather a first-order dipole-dipole interaction. This occurs between molecules in rotational states that differ by exactly one quantum. According to dipole selection rules, the permanent dipole moment couples rotational states $j$ and $j\pm 1$. A pair of molecules in, for example, $j+j'=0+1$, therefore couples to $j'+j=1+0$. These two states are degenerate and thus give rise to a resonant dipole-dipole interaction \cite{mason:62}.

At low temperatures, comparable to energy differences between hyperfine states arising from nuclear and electronic spin, the hyperfine structure within each rotational manifold cannot be neglected. 
Labeling these rotation-hyperfine states with $j,F$,
resonant dipolar interactions occur if $j,F+j',F'$ is coupled by dipole-dipole to $j',F'+j,F$, that is, if the transition $j,F\rightarrow j',F'$ is dipole allowed.
In cases where, for example, $0,F+1,F'$ does \textit{not} couple to $1,F'+0,F$, we find van der Waals interactions that now stem from virtual (de-)excitation to other hyperfine states within the same $j+j'=0+1$ manifold. 
Due to the small excitation energies on the order of the hyperfine splittings, this \textit{hyperfine van der Waals} interaction can be orders of magnitude stronger than rotational van der Waals.
We show in this paper that both attractive and repulsive hyperfine van der Waals interactions can be realized.

\begin{figure}
    \centering
    \includegraphics[width=0.95\linewidth]{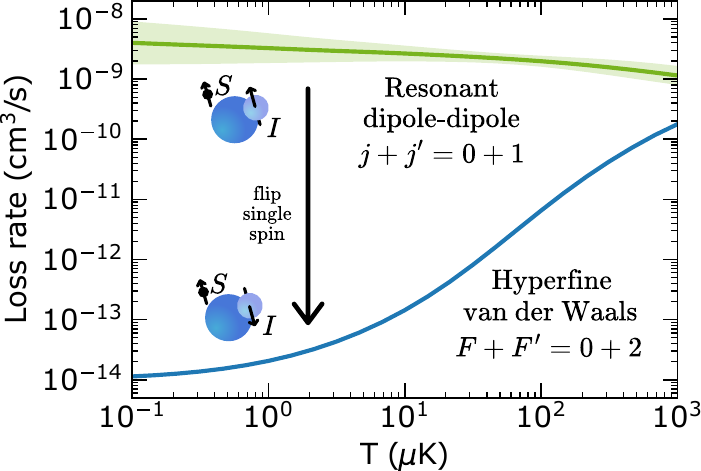}   
    \caption{\textbf{Decreasing CaF collisional loss rates by hyperfine van der Waals repulsion} between molecules in states $j,F=0,0$ and $1,2$ compared to other $j+j'=0+1$ collisions with high losses due to resonant dipole-dipole interactions, with mean and standard deviation given by the green shaded region. By changing spin state, e.g., by merely changing the fluorine nuclear spin state, one can effectively turn off the dipole-dipole interaction, resulting in losses decreased by up to five orders of magnitude.}
    \label{fig:hfvsrdd}
\end{figure}

The repulsive hyperfine van der Waals interactions can suppress two-body collisional loss. We show this suppression with a focus on CaF, which is the prototypical laser-coolable molecule \cite{anderegg:18,cheuk:18ref,caldwell:19}. We also find a universal dependence of these loss rates on the magnitude of the hyperfine splittings compared to the dipolar energy scale. For the closed-shell bialkalis, the hyperfine coupling is orders of magnitude smaller, and suppression of loss does not occur. We find that our results do universally apply to open-shell molecules, and demonstrate this explicitly for MgF, SrF, BaF, and YO. For CaF, we show in Fig.~\ref{fig:hfvsrdd} that two molecules in $j+j'=0+1$ usually experience resonant dipole-dipole interactions, leading to fast losses. However, by carefully selecting initial hyperfine states, e.g., by changing the nuclear spin state, one can effectively turn off the dipole-dipole interaction. The repulsive hyperfine van der Waals interaction can then decrease collisional loss by up to five orders of magnitude.

This paper is organized as follows.
Section \ref{sec:hf_vdw} further details the hyperfine van der Waals interaction with CaF as an example, describing for which states these interactions occur and for which they are attractive or repulsive.
Section \ref{sec:model} provides the necessary theoretical framework and computational details in order to calculate collisional loss rates.
Section \ref{sec:collisions} discusses the interaction potentials and loss rates quantitatively for CaF.
Section \ref{sec:molecules} describes the universality of hyperfine van der Waals repulsion and quantifies decreased loss rates for other molecular systems. 
Section \ref{sec:fields} shows how reduced loss rates persist in the presence of magnetic fields up to 10 G. 
Section \ref{sec:conclusion} concludes this paper.

\section{Hyperfine van der Waals \label{sec:hf_vdw}}

The dominant electrostatic interaction between two neutral polar molecules is the anisotropic dipole-dipole interaction
\begin{equation}
    \hat{V}_\mathrm{dd}(R)=\frac{1}{4\pi\epsilon_0R^3}\left(\hat{\bm{d}}^{(A)}\cdot\hat{\bm{d}}^{(B)}-3\hat{d}_z^{(A)}\hat{d}_z^{(B)}\right)\,,
    \label{eq:Vdd}
\end{equation}
written here in the body-fixed frame where the intermolecular axis lies along the $z$-axis, with dipole moment $\hat{\bm{d}}$ and intermolecular distance $R$. Without external fields, dipolar molecules rotate freely, and when prepared in a pure rotational state, the first-order dipole-dipole interaction vanishes. The only exception here is the interaction between two molecules that are in rotational states $j$ and $j'=j+1$, like $j+j'=0+1$. The pair state $0+1$ is degenerate with the state $1+0$, which couple together due to the dipole selection rule coupling $j$ with $j\pm1$. This results in a resonant dipole-dipole interaction \cite{mason:62}, which is a long-range anisotropic $R^{-3}$ interaction that results in fast losses.

If there is no first-order coupling, the dominant interaction between molecules in some state $|n\rangle\equiv|n_A,n_B\rangle$ arises from second-order dipole-dipole coupling to other states $|n'\rangle\equiv|n'_A,n'_B\rangle$, resulting in a van der Waals interaction
\begin{equation}
    \sum_{n'\neq n}\frac{\left|\langle n'|\hat{V}_\mathrm{dd}(R)|n\rangle\right|^2}{E_n-E_{n'}}=-\frac{C_6}{R^6},
    \label{eq:secondorder}
\end{equation}
with the dipole-dipole operator $\hat{V}_\mathrm{dd}(R)$, the energy $E_n$ of state $|n\rangle$, and the van der Waals coefficient $C_6$. Note that a strong interaction is obtained for molecules in state $|n\rangle$ when there is coupling to a virtual state $|n'\rangle$ that is very close by in energy, i.e., when there is a term in Eq.~\eqref{eq:secondorder} with a small denominator. Moreover, the sign of the interaction depends on the sign of the energy differences, so on whether the dominant virtual state is higher or lower in energy. 

The van der Waals interaction consists of contributions from second-order coupling to other electronic or rotational states. For most molecules, the electronic contribution dominates, generally leading to an attractive interaction. For dipolar molecules, the electronic contribution is typically two orders of magnitude smaller than the rotational one \cite{zuchowski:13}, due to the large differences in excitation energy. We recently found that the small energy differences between rotational states can even result in repulsive rotational van der Waals interactions \cite{walraven:24a}. This occurs when the two molecules are prepared in rotational states that differ by at least two quanta. At low temperatures, we also need to take the hyperfine structure into account. The energy differences between hyperfine states are orders of magnitude smaller than for rotational states. This can result in even stronger second-order interactions. We describe here how such states can be chosen strategically.

From the nuclear spin $I$, electronic spin $S$, and molecular rotation $j$, the total angular momentum $F$ is defined through $\hat{\bm{F}}=\hat{\bm{j}}+\hat{\bm{S}}+\hat{\bm{I}}$, which is used to label each hyperfine state. The hyperfine coupling gives rise to hyperfine splitting of the various spin states. When the collision energy becomes comparable to or smaller than these splittings, the different spin states within a rotational manifold can no longer be treated as being nearly degenerate. 

Interactions between molecules in different spin states $j,F$ depend on the selection rules on $j$ and $F$. More specifically, it depends on whether the transition $j,F\to j',F'$ is allowed, which couples the degenerate states $j,F+j',F'$ and $j',F'+j,F$ by dipole-dipole. If $j'\neq j\pm1$, e.g., $j+j'=0+0$ or $0+2$, rotational van der Waals interactions dominate. This is independent of $F$ as the hyperfine splittings are negligible compared to the rotational splitting. If $j'=j\pm1$, e.g., $j+j'=0+1$, then the interaction between molecules does depend on $F$ and $F'$. The dipole selection rule for the total angular momentum is $\Delta F=0,\pm1$ with $F=0\to F'=0$ forbidden. This means that two states for which $F=F'=0$ or $|F-F'|\geq2$ do not couple. Therefore, two molecules that are prepared in, e.g., $0,F+1,F'$ with total angular momenta that are zero or two or more apart do \textit{not} interact via first-order resonant dipole-dipole.

The main idea here is that we can prepare two molecules in $0,F+1,F'$ with $0,F\to1,F'$ forbidden by the selection rule on $F$. The dominant second-order coupling then takes place to other hyperfine states within $j+j'=0+1$. If the coupling is dominant to molecule-molecule states above the initial state, then an attractive interaction is expected. Otherwise, by coupling to other states below in energy, this results in a repulsive interaction. The energy differences here are in the order of hyperfine splitting, which produces strong van der Waals interactions, reducing collisional loss at low temperatures, as discussed in Sec.~\ref{sec:collisions}.

\begin{figure}
    \centering
    \includegraphics[width=0.95\linewidth]{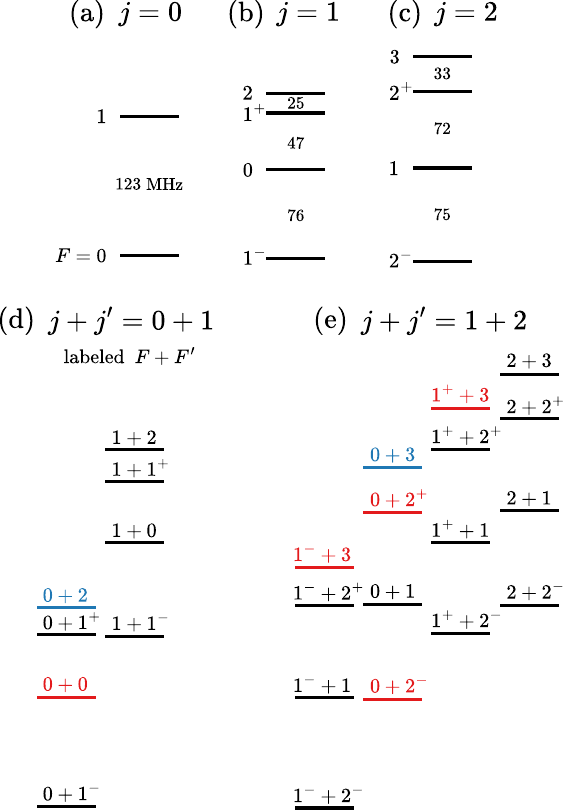}   
    \caption{\textbf{CaF hyperfine structure} labeled with total angular momentum $F$ for the rotational states (a) $j=0$, (b) $j=1$ and (c) $j=2$, as well as the combined pairs (d) $j+j'=0+1$ and (e) $j+j'=1+2$, with states given by $F+F'$. In (a-c), all energy level spacings are given in MHz. In (d) and (e), the $F+F'$ states shown in blue and red do not couple to $F'+F$. The states shown in blue show decreased losses due to repulsive interactions, and those shown in red either couple in first order to other states, or attractively in second order, both resulting in fast losses.}
    \label{fig:hfstructure}
\end{figure}

From here on, we use CaF as our main example to illustrate hyperfine van der Waals interactions, although we expand on universality and other molecules in Sec.~\ref{sec:molecules}. For CaF, the ground rotational state $j=0$ has two hyperfine states, namely $F=0$ and $1$, split by spin-spin coupling. The higher $j$ states have four levels due to additional spin-rotation coupling. For example, the hyperfine states for $j=1$ are given by $F=1^-,0,1^+,2$. Here, we label states of the same $F$ with $-$ and $+$ for those lower or higher in energy. In Figs. \ref{fig:hfstructure}(a)-(c) the hyperfine structure is shown for rotational states $j=0$, 1 and 2 with energy levels split by around 25 to 123 MHz. The hyperfine splittings are determined by diagonalization of the hyperfine Hamiltonian in Eq.~\ref{eq:monH} with coupling parameters as given in Table~\ref{tab:constants}. In order to understand the second-order coupling, it is more insightful to look at the combined energy levels for the pair of molecules, which for $j+j'=0+1$ and $1+2$ are given in Figs. \ref{fig:hfstructure}(d) and \ref{fig:hfstructure}(e), respectively. We discuss these separately below.

As highlighted in color in Fig. \ref{fig:hfstructure}(d) for $j+j'=0+1$, there exist two molecule-molecule states that do not interact through resonant dipole-dipole, but through hyperfine van der Waals. These are $0,F'+1,F'=0,0+1,2$ and $0,0+1,0$, shown in blue and red, respectively, since $0,0\to1,2$ and $0,0\to1,0$ are not dipole allowed. Whether the hyperfine van der Waals interaction in these states is attractive or repulsive depends on which other states they predominantly couple to. For $F+F'=0+0$, this is the $F+F'=1+1^-$ state, which is located higher in energy, resulting in an attractive van der Waals interaction. For $F+F'=0+2$, the main coupling also occurs with $F+F'=1+1^-$, which this time is located below in energy. This means that the hyperfine van der Waals interaction for $F+F'=0+2$ is repulsive and that this state is the primary candidate within $j+j'=0+1$ to find strongly suppressed loss rates.

Similar states can also be found within $j+j'=1+2$ and so on, although the hyperfine structure is more involved, as seen in Fig.~\ref{fig:hfstructure}(e). Note that the number of hyperfine levels and the overall hyperfine structure remain the same for higher rotational states, as shown for $j+j'=2+3$ in the Appendix. For $j+j'=1+2$, the main candidate for a repulsive interaction is the $F+F'=0+3$ state, shown in blue. The states that are shown in red either result in attractive van der Waals interactions, or are accidentally nearly degenerate with another state it couples to, still resulting in resonant dipole-dipole interactions. In general, for CaF, we find the most repulsive interactions between molecules in $j,F+(j+1),{F'}$ when $F=j-1$ and $F'=j+2$. This state can only couple to one other state, namely $F+F'=j'+j$, which is lower in energy. 

\section{Theoretical framework and computational details \label{sec:model}}

The monomer Hamiltonian for a $^2\Sigma$ molecule $X$, treated as a rigid rotor with a single nuclear spin $1/2$, is given by
\begin{equation}
    \hat{H}^{(X)}=B\hat{j}^2+c_1\hat{\bm{S}}\cdot\hat{\bm{j}}+c_2\hat{\bm{I}}\cdot\hat{\bm{j}}+c_3\hat{H}_3+c_4\hat{\bm{S}}\cdot\hat{\bm{I}}\,,
    \label{eq:monH}
\end{equation}
where the terms respectively describe rotational kinetic energy, electron spin-rotation coupling, nuclear spin-rotation coupling, and the tensor and scalar spin-spin coupling \cite{childs:81,brown:03}. The tensor term can be written in spherical tensor form as
\begin{equation}
    \hat{H}_3=-\sqrt{30}\left[ \left[ \hat{\bm{S}} \otimes \hat{\bm{I}} \right]^{(2)} \otimes C_{2}\left(\hat{R}\right) \right]^{(0)}_0\,,
\end{equation}
with $C_2(\hat{R})$ the rank-2 tensor whose spherical components are the Racah normalized spherical harmonics, and
\begin{align}
    \left[\hat{A}_{k_A} \otimes \hat{B}_{k_B}\right]^{(k)}_q = \sum_{q_A,q_B} \hat{A}_{k_A, q_A} \hat{B}_{k_B,q_B} \langle k_A q_A k_B q_B | k q\rangle
\end{align}
is the $q$ spherical component of the rank $k$ irreducible tensor product of the rank $k_A$ spherical tensor $\hat{A}$ and the rank $k_B$ spherical tensor $\hat{B}$. The values for the molecular constants used throughout the calculations can be found in Table~\ref{tab:constants}.

For the scattering of two molecules with monomer Hamiltonians $\hat{H}^{(A)}$ and $\hat{H}^{(B)}$ given by Eq.~\eqref{eq:monH}, we define the total dimer Hamiltonian
\begin{equation}
    \hat{H}=-\frac{\hbar^2}{2\mu}\frac{1}{R}\frac{\mathrm{d}^2}{\mathrm{d}R^2}R+\frac{\hat{L}^2}{2\mu R^2}+\hat{H}^{(A)}+\hat{H}^{(B)}+\hat{V}_\mathrm{dd}(R)\,,
    \label{eq:dimH}
\end{equation}
with reduced mass $\mu$, intermolecular distance $R$, end-over-end rotational angular momentum $\hat{\bm{L}}$ and the dipole-dipole interaction
\begin{align}
    &\hat{V}_\mathrm{dd}(R) = -\frac{\sqrt{30} d^2}{4\pi\epsilon_0 R^3} \nonumber\\
    &\hspace{5mm}\times\bigg[ \Big[ C_{1}\left(\hat{r}^{(A)} \right) \otimes C_{1}\left(\hat{r}^{(B)} \right)\Big]^{(2)}\otimes C_{2}\left(\hat{R}\right) \bigg]^{(0)}_0\,,
\end{align}
with molecular axes $\hat{r}^{(A)}$ and $\hat{r}^{(B)}$ and dipole moment $d$.

Throughout our calculations, we use an uncoupled basis set consisting of angular momentum states for the molecular rotation together with electronic and nuclear spin states, yielding the uncoupled monomer states $|n^{(X)}\rangle=|j,m\rangle|S,M_S\rangle|I,M_I\rangle$. Together with the end-over-end rotation $\hat{\bm{L}}$ this yields the dimer basis functions $|n^{(A)}\rangle|n^{(B)}\rangle|L,M_L\rangle$. To reduce the basis set size and consequently computational cost, we perform separate calculations for different $M_\mathrm{tot}=m^{(A)}+M_S^{(A)}+M_I^{(A)}+m^{(B)}+M_S^{(B)}+M_I^{(B)}+M_L$ and different parities $(-1)^{j_A+j_B+L}$, and adapt the basis functions to permutation symmetry. Our results are converged up to a few percent with $j_\mathrm{max}=1$, $L_\mathrm{max}=10$ and $|M_{L,\mathrm{in}}|_\mathrm{max}=5$. Results in Fig.~\ref{fig:lossrates} involving $j+j'=1+2$ are converged to similar precision using $j_\mathrm{max}=2$ and for $j+j'=0+2$ using $j_\mathrm{max}=3$. Results in Fig.~\ref{fig:lossrates_scaling} are converged using $j_\mathrm{max}=2$.

At zero (or low) fields, the molecular eigenstates are represented more closely by the coupled Hund's case (b) representation $|(j,S),J,I;F,M_F\rangle$ with $\hat{\bm{J}}=\hat{\bm{j}}+\hat{\bm{S}}$ and $\hat{\bm{F}}=\hat{\bm{J}}+\hat{\bm{I}}$, which can be obtained from the uncoupled states via
\begin{align}
    |(j,&S),J,I;F,M_F\rangle = \sum_{m,M_S,M_I}\langle j,m,S,M_S|J,m+M_S\rangle\nonumber\\
    &\times \langle J,m+M_S,I,M_I|F,M_F\rangle|j,m\rangle|S,M_S\rangle|I,M_I\rangle\,.
\end{align}
In this paper, we typically denote those monomer states for brevity by $j,F$ and the dimer states by noting $j+j'$, $F+F'$, or $j,F+j',F'$.

With the Hamiltonian and basis sets now discussed, we turn to collisions. The collisional loss rates are obtained through coupled-channels scattering calculations using the renormalized Numerov method \cite{johnson:78ref}. The short-range losses are modeled by an absorbing boundary condition, which assumes 100\% loss of flux going into the short-range reactive channels at the distance where this boundary condition is imposed. This matches experimentally observed losses for ultracold molecules \cite{takekoshi:14,voges:20,guo:18,ye:18,gregory:19,cheuk:20ref}. At large distances, we impose $S$-matrix boundary conditions \cite{janssen:13ref}. With these boundary conditions, results are converged with a radial grid ranging from 50 to 100\,000 $a_0$ with at least 30 points per local de Broglie wavelength. The resulting $S$-matrix after propagation contains the state-to-state scattering probabilities from which we get collisional cross sections, including those pertaining to collisional loss. There are two loss mechanisms: short-range loss through the absorbing boundary, and inelastic loss to other channels \cite{walraven:24a}. Inelastic collisions that change $M_F$ within the initial $F$ state are not considered as loss, as these states are degenerate at infinite separation and zero field. To obtain thermal rate coefficients, these energy-dependent cross sections are then averaged over a Maxwell-Boltzmann distribution, where we perform the integration using 37 logarithmically spaced collision energies between $10^{-2}$ and $10^{4}$~$\mu$K.

\section{Interaction potentials and loss rates \label{sec:collisions}}

\begin{figure}
    \centering
    \includegraphics[width=0.95\linewidth]{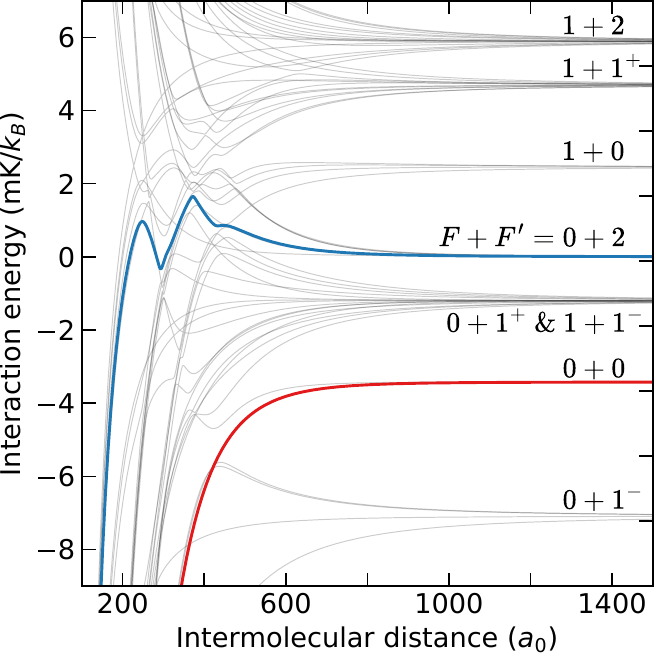}   
    \caption{\textbf{CaF interaction potentials} for all hyperfine pairs within $j+j'=0+1$. Only $M_\mathrm{tot}=0$ with $L=0$ and $2$ are shown here. Highlighted in blue and red are the $L=0$ adiabats of the repulsive $F+F'=0+2$ pair and the attractive $F+F'=0+0$ pair.}
    \label{fig:potentials}
\end{figure}

We obtain adiabatic interaction potentials by diagonalizing the dimer Hamiltonian \eqref{eq:dimH} at each distance $R$, excluding the radial kinetic energy term. The resulting potentials for all hyperfine states within $j+j'=0+1$ are given in Fig.~\ref{fig:potentials}. We highlight the curves shown in blue and red, which represent the lowest adiabats connecting to the s-wave $F+F'=0+2$ and $F+F'=0+0$ channels, respectively. As discussed in Section \ref{sec:hf_vdw}, the second-order coupling to $F+F'=1+1^-$ results in a repulsive $R^{-6}$ potential in the long range for $F+F'=0+2$ and an attractive one for $F+F'=0+0$. At shorter distances, the perturbative argument breaks down, and the molecules polarize each other, resulting in a highly attractive interaction. For $F+F'=0+2$, a barrier is formed that can ``shield'' collisions with collision energies below the barrier height, which is on the order of mK. Residual losses then stem from tunneling through this barrier, after which the molecules are lost in the short range, as well as from inelastic coupling to other hyperfine states.

One other state also requires attention, the $F+F'=0+1^+$ state. Although it has resonant first-order dipole-dipole interactions, it also couples to the $1+1^-$ state, which lies closely below by only 0.36~MHz $\approx$ 16~$\mu$K. This results in a repulsive potential that can decrease losses, but only when the collision energy is small compared to this 16~$\mu$K \footnote{The coincidental near degeneracy of $F+F'=0+1^+$ and $1+1^-$ is sensitive to the precise hyperfine constants.}.

\begin{figure}
    \centering
    \includegraphics[width=0.95\linewidth]{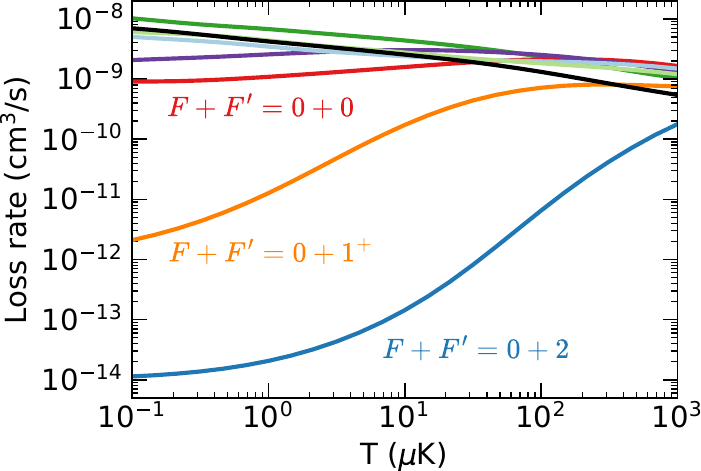}   
    \caption{\textbf{Temperature dependence of CaF loss rates for all hyperfine states within $j+j'=0+1$.} The lowest loss rates are found for $F+F'=0+2$ (blue), resulting from hyperfine van der Waals repulsion. At lower temperatures, also $F+F'=0+1^+$ (orange) shows decreased loss rates. The $F+F'=0+0$ state (red) has an attractive van der Waals interaction and shows universal loss. All other hyperfine states not labeled show high loss rates due to resonant dipolar interactions.}
    \label{fig:lossrates01}
\end{figure}

The distinct interactions among all hyperfine pairs lead to vastly different loss rates. For each hyperfine state of CaF within $j+j'=0+1$, these rates are shown as a function of temperature in Fig.~\ref{fig:lossrates01}. The various hyperfine states with the highest loss rates correspond to resonant dipole-dipole losses. The $F+F'=0+0$ state highlighted in red reaches slightly lower losses at low temperatures corresponding to universal loss due to its attractive $R^{-6}$ interaction \cite{idziaszek:10ref}. The $0+1^+$ state shown in orange shows decreased loss rates at temperatures that are roughly lower than 16~$\mu$K. Finally, the state that has been of most interest so far, the $0+2$ state presented in blue, shows a loss rate that becomes as small as $10^{-14}$~cm$^3$/s due to the potential barrier, which height is on the order of mK.

\begin{figure}
    \centering
    \includegraphics[width=0.95\linewidth]{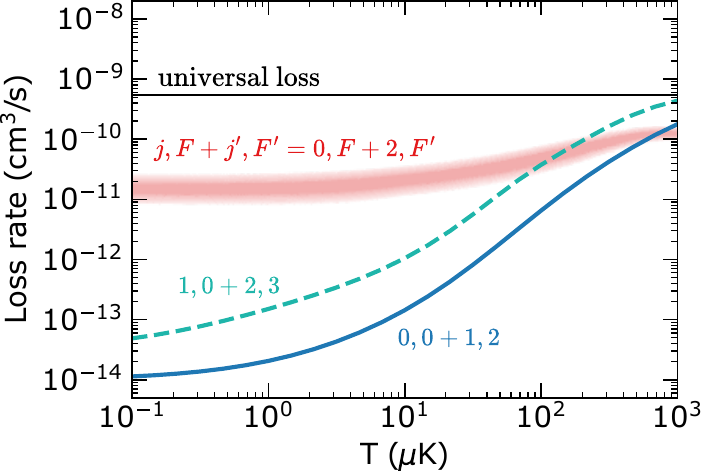}   
    \caption{\textbf{Temperature dependence of CaF loss rates for various rotational states.} The horizontal black line represents universal loss \cite{idziaszek:10ref}, expected when both molecules are in the same rotational state. The shaded area in red represents loss rates for CaF in various hyperfine states $0,F'+2,{F'}$, resulting in repulsive rotational van der Waals interactions \cite{walraven:24a}. The solid blue and dashed cyan lines represent hyperfine van der Waals loss for, respectively, $0,0+1,2$ and $1,0+2,3$.}
    \label{fig:lossrates}
\end{figure}

The two-body loss that ground state ultracold molecules typically suffer from is often close to universal loss \cite{idziaszek:10ref}, which for CaF is $5.4\times10^{-10}$~cm$^3$/s. When comparing loss from collisions for $F+F'=0+2$ to this universal loss rate, hyperfine van der Waals repulsion can lower collisional loss by up to five orders of magnitude, as shown in Fig.~\ref{fig:lossrates}. We also show in this figure that this reduction of the loss rate is not limited to $j+j'=0+1$, but also occurs for $j+j'=1+2$. In the latter case, the lowest loss rates are found for $F+F'=0+3$ as shown by the dashed cyan line. For higher rotational states, we generally find that the energy difference to the state it couples to increases. 
This results in loss rates that are suppressed by orders of magnitude below the universal rate,
although the suppression is less effective than for $j+j'=0+1$.

In Fig.~\ref{fig:lossrates}, we also compare loss rates of molecules in $j+j'=0+2$, which exhibit repulsive rotational van der Waals interactions, to the loss rates due to hyperfine van der Waals repulsion. The shaded region for $j+j'=0+2$ indicates the spread of loss rates for various hyperfine states within this rotational manifold, which shows the weak sensitivity of hyperfine structure for those collision partners. Hyperfine van der Waals repulsion can suppress loss rates by an additional three orders of magnitude.

\section{Universality and other diatomic molecules \label{sec:molecules}}

Hyperfine van der Waals interactions are not unique to CaF. We find such interactions in even the simplest systems with a single spin $S=1/2$. Molecules can then be prepared in, e.g., $1,{1/2}+2,{5/2}$, where $\Delta F>1$ does not allow coupling to $2,{5/2}+1,{1/2}$. The dominant interaction then results from second-order dipole-dipole coupling to other $F+F'$ states within $j+j'=1+2$. Whether the resulting hyperfine van der Waals interaction are in general attractive or repulsive then depends on the specific molecular hyperfine structure.

\begin{figure}
    \centering
    \includegraphics[width=0.95\linewidth]{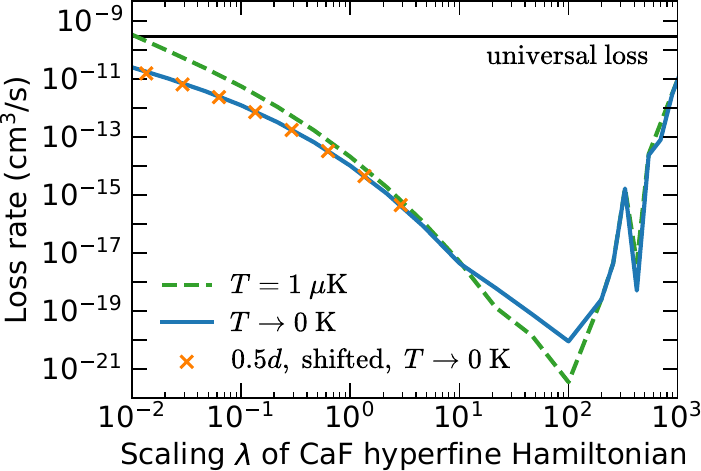}   
    \caption{\textbf{Universality of loss rates} upon scaling of the CaF hyperfine Hamiltonian by scaling hyperfine parameters in Eq. \eqref{eq:monH} by a factor $\lambda$. Loss rates decrease with larger energy spacing between hyperfine levels. At scaling by more than $\lambda=10^2$, the hyperfine energy levels start to cross those of $j+j'=1+2$. Halving the dipole moment and shifting the loss rate curve as described in the main text coincides with the loss rates for $T\to0$, demonstrating universal scaling laws. Typical hyperfine parameters for bialkali would be found below $\lambda=10^{-3}$ scaling and therefore show no significant hyperfine van der Waals repulsion. }
    \label{fig:lossrates_scaling}
\end{figure}

Hyperfine van der Waals repulsion depends on the magnitude of the hyperfine energy scale compared to the dipolar energy scale. We show this in Fig.~\ref{fig:lossrates_scaling}, where we artificially scale all the hyperfine terms in the CaF Hamiltonian $\eqref{eq:monH}$ by an overall scaling parameter $\lambda$ such that relative splittings are kept identical. At larger overall hyperfine coupling, the separation between energy levels increases. Since the repulsive interaction potential forms a barrier due to coupling with higher-lying levels, the barrier height increases as well. This effect therefore decreases loss rates at larger hyperfine coupling. Scaling the hyperfine couplings up by more than a factor of $\lambda=10^2$ causes the hyperfine energy levels of $j+j'=0+1$ to cross those of the $j+j'=1+2$ manifold. This shows that for hyperfine couplings smaller than the rotational constant, which is typically the case, larger hyperfine splittings lead to lower loss rates. Scaling down the hyperfine couplings leads to loss rates similar to resonant dipole-dipole collisions, as if there were no hyperfine.

We also find universality of these loss rates in dipole-dipole units. The natural dipolar length is given by $a_\mathrm{dd}=2\mu C_3/\hbar^2$ and the dipolar energy by $E_\mathrm{dd}=\hbar^2/(2\mu a_\mathrm{dd}^2)$. In the limit $T\to0$, the only energy and length scales are those of the hyperfine and dipole-dipole interactions, where temperature and the de Broglie wavelength do not determine additional energy and length scales. The loss rates then depend solely on the ratio $E_\mathrm{hf}/E_\mathrm{dd}$, with $E_\mathrm{hf}$ the natural hyperfine energy. This means that for the loss rate $k$ we find that $k/a_\mathrm{dd}$ is a universal function of $\lambda/E_\mathrm{dd}$. Halving the dipole moment then increases $\lambda/E_\mathrm{dd}$ by a factor of 16 and increases $k/a_\mathrm{dd}$ by a factor of 4. This behavior is illustrated in Fig.~\ref{fig:lossrates_scaling} by the orange cross marks, where the curve has been shifted to lower $\lambda$ by a factor of 16 and loss rates have been scaled back up by a factor of 4. This rescaled curve aligns perfectly with the $T\to0$ results with the original dipole moment, demonstrating universality. This scaling behavior also explains why suppressed loss rates are not found for the bialkali, even though there exist transitions $j,F\to j',F'$ that are not dipole allowed. Hyperfine states of bialkalis are split by tens of kHz, compared to tens of MHz hyperfine splittings for CaF. This corresponds to a scaling factor $\lambda$ below $10^{-3}$, for which collisional loss is not suppressed. Therefore, molecules with hyperfine coupling arising from electron spin are the most promising candidates for observing decreased loss rates via hyperfine van der Waals repulsion.

\begin{figure}
    \centering
    \includegraphics[width=0.95\linewidth]{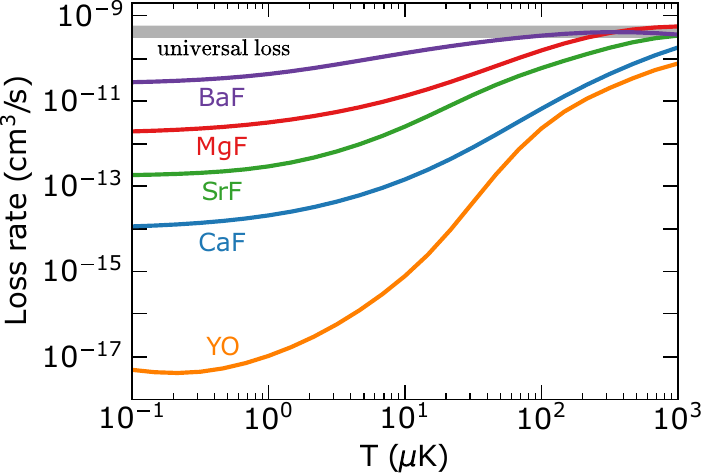}   
    \caption{\textbf{Temperature dependence of loss rates for various molecules in states with hyperfine van der Waals repulsion.} For the series MgF, CaF, SrF and BaF, loss rates are shown for the $j,F+j',F'=0,0+1,2$ state and for YO the $0,0+1,0$ state. The horizontal gray bar represents the range of universal loss for all shown molecules.}
    \label{fig:lossrates_others}
\end{figure}

We investigated several candidate molecules and found that MgF, CaF, SrF, BaF and YO all show reduced loss rates due to hyperfine van der Waals repulsion. The values of molecular constants used can be found in Table~\ref{tab:constants} and the hyperfine structures for $j+j'=0+1$ in the Appendix. The resulting loss rates as a function of temperature are shown in Fig.~\ref{fig:lossrates_others}. All molecules show a similar trend of loss rates decreasing at lower temperatures, though all show various orders of magnitude suppression.

For the earth-alkaline monofluoride series, the hyperfine structure is rather similar, though the relative location of the energy levels does determine differences between them. At low temperatures, we find that loss rates increase as CaF $<$ SrF $<$ MgF $<$ BaF. For SrF and BaF, the main source of loss is short-range loss. As the coupling of the electron spin to the nuclear spin decreases (see $c_4$ in Table~\ref{tab:constants}) and coupling to rotation increases ($c_1$ in Table~\ref{tab:constants}), the gap between the $F+F'=0+2$ and the $1+1^-$ state below increases, while the one to the $1+1^+$ state above decreases. This means that the repulsive barrier decreases from MgF to BaF. This increases short-range loss, which becomes dominant for SrF and BaF, hence why higher loss rates are expected than for CaF. For MgF and CaF, the residual loss is dominated by inelastic loss to other hyperfine states, which is relatively larger for MgF. The trade-off between short-range and inelastic losses results in the lowest loss rates for CaF.

For YO, the hyperfine structure is inverted compared to CaF. It does have the same quanta for the spin states, i.e., $S=I=1/2$, but all coupling parameters have opposite sign. This means that for YO, the interaction for molecules in $F+F'=0+2$ is attractive, while for $0+0$ it is repulsive. Potential energy curves for $j+j'=0+1$ can be found in the Appendix. Loss rates for $F+F'=0+0$ are the ones shown in Fig.~\ref{fig:lossrates_others}. A unique feature of this state is that there is no higher-lying state in $j+j'=0+1$ that also couples to $F+F'=0+0$. This greatly reduces short-range loss to the point where only inelastic loss dominates, yielding loss rates as low as $10^{-17}$ cm$^3$/s.

\begin{table*}
    \centering
    \caption{Molecular constants used in this work. Each molecule has a $^2\Sigma$ electronic ground state with $S=1/2$ and $I=1/2$.}
    \begin{tabular}{lrcrcrcrcrc}
        \hline\hline
        Constant & $^{24}$Mg$^{19}$F & Ref. & $^{40}$Ca$^{19}$F & Ref. & $^{88}$Sr$^{19}$F & Ref. & $^{138}$Ba$^{19}$F & Ref. & $^{89}$Y$^{16}$O & Ref. \\\hline
        $B$ (GHz)   & 15.50 & \cite{anderson:94}    & 10.27  & \cite{childs:81} & 7.511  & \cite{domaille:77} & 6.470 & \cite{jenkins:32} & 11.63   & \cite{weltner:67} \\
        $d$ \ (D)     & 2.88  & \cite{doppelbauer:22} & 3.07   & \cite{childs:84} & 3.496  & \cite{kandler:89}  & 3.17  & \cite{ernst:86}   & 4.52    & \cite{steimle:90} \\
        $c_1$ (MHz) & 50.70 & \cite{anderson:94}    & 39.659 & \cite{childs:81} & 75.022 & \cite{childs:81a}  & 80.92 & \cite{ernst:86}   & -9.225  & \cite{childs:88}  \\
        $c_2$ (kHz) &       &                       & 29.07  & \cite{childs:81} & 2.3    & \cite{childs:81a}  &       &                   & -2.6    & \cite{childs:88}  \\
        $c_3$ (MHz) & 59.5  & \cite{anderson:94}    & 13.373 & \cite{childs:81} & 10.09  & \cite{childs:81a}  & 2.74  & \cite{ernst:86}   & -9.413  & \cite{childs:88}  \\
        $c_4$ (MHz) & 214.2 & \cite{anderson:94}    & 122.57 & \cite{childs:81} & 107.17 & \cite{childs:81a}  & 83.66 & \cite{ernst:86}   & -772.39 & \cite{childs:88}  \\
        \hline\hline
    \end{tabular}
    \label{tab:constants}
\end{table*}

\section{Dependence on magnetic fields \label{sec:fields}}

In zero field, each $F$ state has $2F+1$ degenerate $M_F$ components. So far, we have assumed $M_F$-changing collisions within $F$ do not constitute loss. Magnetic fields, however, split up these $M_F$ states by the Zeeman effect. Collisions that change $M_F$ could then lead to energy release and therefore loss or heating, depending on the trap depth. We therefore explore for CaF up to what magnetic field strength collisional loss is still suppressed, where we now consider $M_F$-changing collisions to constitute loss due to the energy released by Zeeman relaxation.

The interaction of the electron spin $\bm{S}$ with a magnetic field $\bm{B}$ is given by the Zeeman Hamiltonian
\begin{equation}
    \hat{H}_\mathrm{Zeeman}=-g_e\mu_B\hat{\bm{S}}\cdot\bm{B}\,,
    \label{eq:zeeman}
\end{equation}
with electronic $g$-factor $g_e\approx2.002319$ and bohr magneton $\mu_B=1/2$ in atomic units. The coupling of the magnetic field with the nuclear spin or the rotation of the molecule is neglected, as well as the anisotropic coupling with the electron spin. These further contributions are at least three orders of magnitude smaller than the isotropic electron spin interaction we include here \cite{caldwell:20}. We add the Zeeman Hamiltonian of Eq.~\eqref{eq:zeeman} to the monomer Hamiltonian \eqref{eq:monH}.

\begin{figure}
    \centering
    \includegraphics[width=0.95\linewidth]{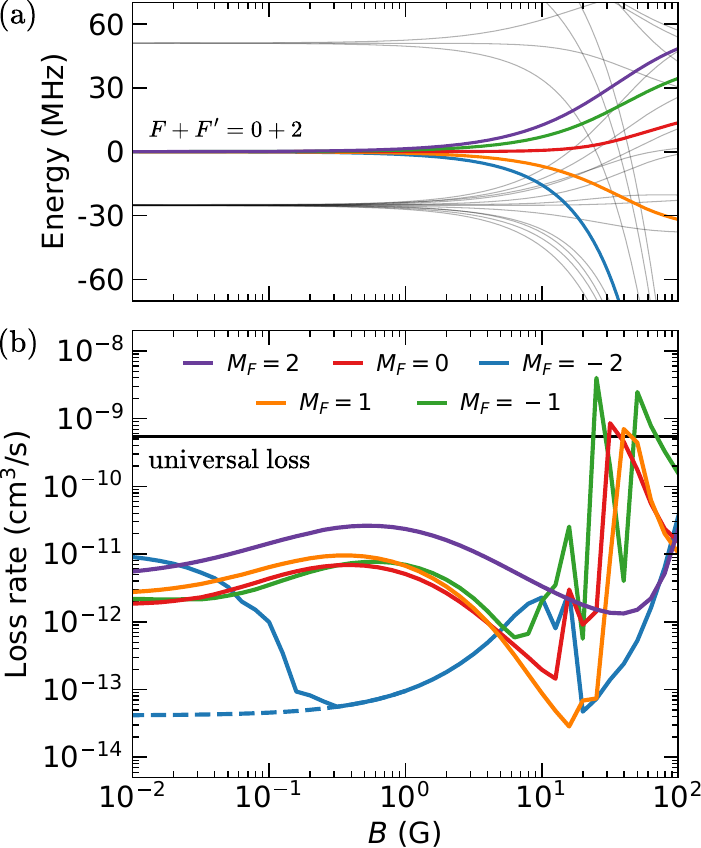}
    \caption{\textbf{Magnetic field dependence.} (a) Energy levels of both molecules at infinite separation as a function of magnetic field $B$. (b) Loss rates as a function of magnetic field at $T=1$~$\mu$K. The different curves correspond to the $F=2$ molecule prepared in its different $M_F$ components. For $M_F=-2$, the additional dashed curve shows loss rates that exclude excitation to other $M_F$ states at low $B$.}
    \label{fig:Bfield}
\end{figure}

For the $F+F'=0+2$ state, the molecule in $F=2$ has components $M_F=-2,-1,0,1,2$, with $M_F=-2$ the lowest in energy in a magnetic field. For CaF, the energy levels at infinite separation as a function of magnetic field are given in Fig.~\ref{fig:Bfield}(a), where it can be seen that states start to cross above magnetic fields of around 10~G. The loss rates of these states at $T=1$~$\mu$K are shown in Fig.~\ref{fig:Bfield}(b). At small $B$, the inelastic loss to other $F+F'$ states is around $10^{-13}$ cm$^3$/s, as shown by the dashed curve. At large $B$, where $F+F'$ levels cross, the loss rate increases and is dominated by transitions to other $F+F'$ states. For fields below 10~G, it is the lowest $M_F=-2$ state that shows the lowest loss rates. For molecules in other $M_F$ states, the dominant loss is Zeeman relaxation to other $M_F$ states within $F+F'=0+2$, with energy released due to the Zeeman splitting. Below 0.1~G, this splitting becomes smaller than the collision energy, where excitation to other $M_F$ becomes energetically available. This also means that the energy release has become small enough that it might not necessarily lead to heating. In general, we find that the suppression of loss rates due to hyperfine van der Waals repulsion should be observable for finite magnetic fields below 10~G.

\section{Conclusion \label{sec:conclusion}}

We found a new interaction, the hyperfine van der Waals interaction, that occurs between molecules in specific combinations of hyperfine states when the rotational quanta of the molecules differ by one. Where molecules in these rotational states typically experience resonant dipole-dipole interactions, selection rules on hyperfine levels prevent certain combinations of spin states from interacting resonantly. This results in hyperfine van der Waals interactions that can either be attractive or repulsive depending on the relative positions of hyperfine energy levels. Our coupled-channels calculations show that the repulsive interaction can lead to loss rates reduced by up to five orders of magnitude below universal loss for CaF, without the need of any external fields. We also found a universal dependence on the ratio between the hyperfine and dipolar energy scales. The hyperfine van der Waals repulsion was shown to be present for the earth-alkaline monofluoride series MgF, CaF, SrF and BaF, as well as for YO. The lowest loss rates were found for YO with eight orders of magnitude suppression compared to universal loss.

Our predictions for the strong suppression of collisional loss can be directly tested experimentally by controlled collisions between molecules prepared in the different hyperfine states. A well-suited testbed would be to merge optical tweezers containing single molecules. Polar molecules, and in particular CaF, have already been loaded individually in optical tweezers \cite{anderegg:19,burchesky:21,zhang:22}, their rotational and hyperfine states have been controlled \cite{cheuk:20ref,ruttley:24}, and tweezers have been merged to study the collision dynamics \cite{cheuk:20ref,anderegg:21}. Loss rates can then be obtained by preparing molecules in the desired hyperfine states prior to merging. A single flip of a nuclear spin then switches off resonant dipolar interactions and suppresses loss by five orders of magnitude.

\appendix*
\section{HYPERFINE STRUCTURE}

\begin{figure}
    \centering
    \includegraphics[width=0.5\linewidth]{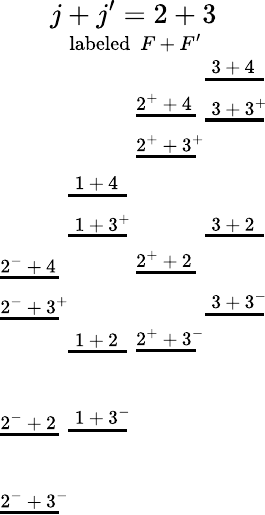}
    \caption{\textbf{Hyperfine structure of CaF} for the combined pair $j+j'=2+3$.}
    \label{fig:app_CaF23}
\end{figure}

The CaF hyperfine energy level structure is very similar for any $j\ge 1$, see Fig.~\ref{fig:hfstructure}; the number of hyperfine states and their positions are the same, while the splitting due to spin-rotation coupling gradually increases with $j$.
As a result, the hyperfine energy levels for a pair of CaF molecules in $j+(j+1)$ with $j\geq1$ are also very similar for different $j$. In Fig.~\ref{fig:app_CaF23}, we give the CaF hyperfine structure for the combined pair of $j+j'=2+3$. The similarities are clear when comparing to $j+j'=1+2$ in Fig.~\ref{fig:hfstructure}(e), which shows that the relative positions can change slightly, while the overall structure remains. The same principle holds for higher $j$.
In general, for CaF, we find the most repulsive interactions between molecules in $j,F+(j+1),{F'}$ when $F=j-1$ and $F'=j+2$.

\begin{figure*}
    \centering
    \includegraphics[width=0.8\linewidth]{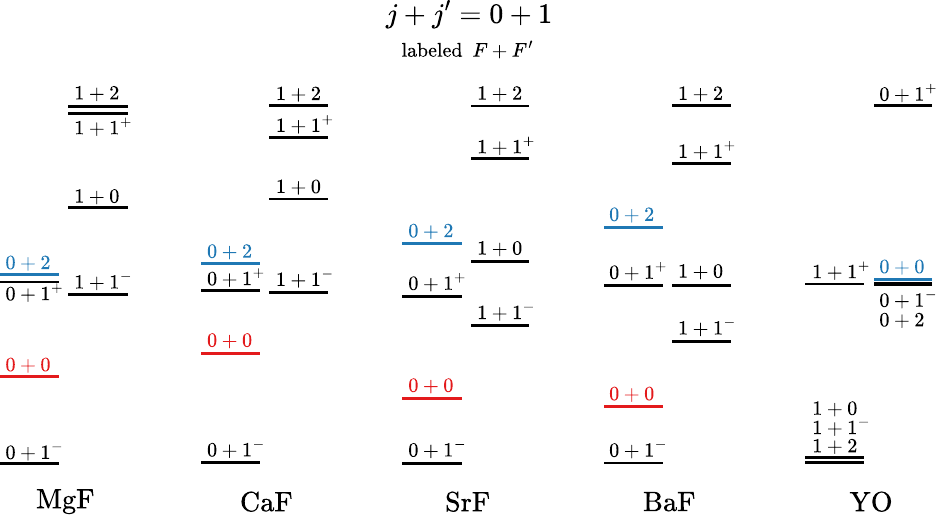}
    \caption{\textbf{Hyperfine structure of MgF, CaF, SrF, BaF, YO} for the combined pair $j+j'=0+1$. Each set of energy levels is scaled to its maximum hyperfine splitting. For the five molecules, these maximum energy differences are, respectively, 453, 271, 278, 247 and 1549 MHz.}
    \label{fig:app_HF01}
\end{figure*}

In Sec.~\ref{sec:molecules} we show loss rates for MgF, CaF, SrF, BaF and YO, of which we provide here the hyperfine level structure for $j+j'=0+1$ in Fig.~\ref{fig:app_HF01}. Most importantly for the earth-alkaline monofluoride series, we find different relative positions of $F+F'=0+2$ to the states to which it couples, $1+1^-$ and $1+1^+$. For YO, the structure is inverted compared to the others, with more dominant spin-spin coupling. The repulsive interaction arises here for $F+F'=0+0$, which lies closely above $1+1^+$ to which it couples. 

\begin{figure}
    \centering
    \includegraphics[width=0.95\linewidth]{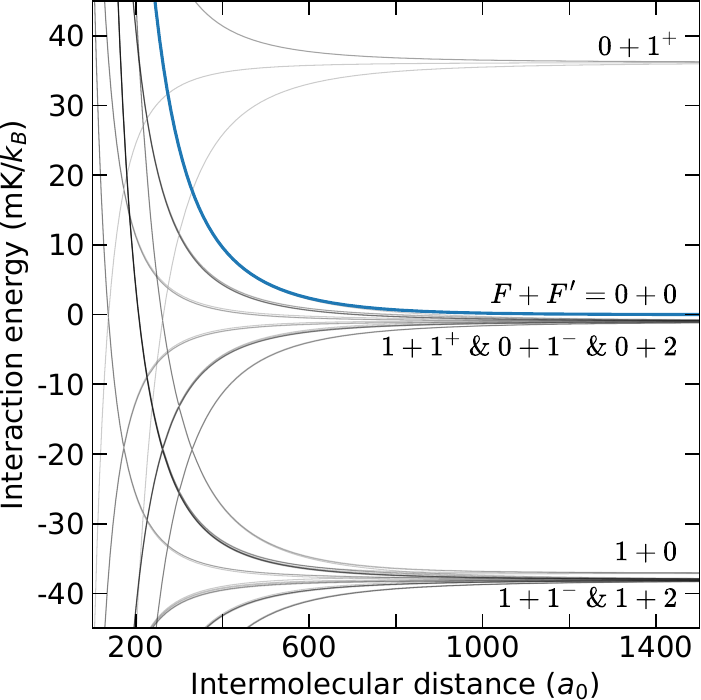}
    \caption{\textbf{Interaction potentials for YO} for all hyperfine states within $j+j'=0+1$. Only $M_\mathrm{tot}=0$ with $L=0$ and $2$ are shown here. Highlighted in blue is the $L=0$ adiabat of the repulsive $F+F'=0+0$ pair.}
    \label{fig:app_YOpot}
\end{figure}

For YO, we also show interaction potentials in Fig.~\ref{fig:app_YOpot}. Since there is no other hyperfine state within $j+j'=0+1$ above $F+F'=0+0$ that couples to that state, a strong repulsive interaction is found, which results in lower loss rates compared to the earth-alkaline monofluoride series.

\end{document}